\documentclass[a4paper,fleqn,usenatbib]{mnras}

\usepackage[T1]{fontenc}
\usepackage{ae,aecompl}

\usepackage{graphicx}	
\usepackage{amsmath}	
\usepackage{amssymb}	

\title[The origin of the occurrence rate profile of gas giants inside 100 days]{The origin of the occurrence rate profile of gas giants inside 100 days}
\author[Ali-Dib et al.]{
Mohamad Ali-Dib$^{1}$$^{,}$$^{2}$\thanks{E-mail: m.alidib@utoronto.ca}, Anders Johansen$^{3}$, and Chelsea X. Huang$^{1}$$^{,}$$^{4}$$^{,}$$^{5}$
\\
$^{1}$Centre for Planetary Sciences, Department of Physical \& Environmental Sciences, University of Toronto at Scarborough,\\
Toronto, ON M1C 1A4, Canada\\
$^{2}$Canadian Institute for Theoretical Astrophysics, 60 St. George St, University of Toronto, Toronto, ON M5S 3H8, Canada\\
$^{3}$Lund Observatory, Department of Astronomy and Theoretical Physics, Lund University, Box 43, 22100 Lund, Sweden\\
$^{4}$Dunlap Institute, University of Toronto, Toronto, ON, M5S3H4, Canada\\
$^{5}$Juan Carlos Torres Fellow, MIT Kavli Institution, Cambridge, MA, 02139
}

\date{Accepted XXX. Received YYY; in original form ZZZ}

\pubyear{2017}

\begin{document}
\label{firstpage}
\pagerange{\pageref{firstpage}--\pageref{lastpage}}
\maketitle


\begin{abstract}
We investigate the origin of the period distribution of giant planets. We try to fit the bias-corrected distribution of giant planets inside 300 days found by \cite{santerne2016} using a planet formation model based on pebble accretion. We investigate two possible initial conditions: a linear distribution of planetary seeds, and seeds injected exclusively on the water and CO icelines. Our simulations exclude the linear initial distribution of seeds with a high degree of confidence. Our bimodal model based on snowlines give a more reasonable fit to the data, with the discrepancies reducing significantly if we assume the water snowline to be a factor 3-10 less efficient at producing planetary seeds. This model moreover performs better on both the warm/hot Jupiters ratio and the Gaussian mixture model as comparison criteria. Our results hint that the giant exoplanets population inside 300 days is more compatible with planets forming preferentially at special locations. 
\end{abstract}

\begin{keywords}
planets and satellites: formation -- planets and satellites: gaseous planets
\end{keywords}


\section{Introduction}
Variations in the period distribution of giant planets can provide a wealth of informations on planet formation scenario. Classical planet formation models predict that giant planets should be more abundant outside of the snowline due to higher isolation masses caused by higher solids density \citep{pollack1996}. Planets, however, can interact with the disk via torques exerted by the spiral arms induced by the planet, and these can push the planet significantly in either radial direction \citep{kley2012}. Moreover, a giant planet can interact with other giant planets or stellar companions, possibly scattering the planet off. These processes are sensitive to the disk's thermal and density structure, and the presence and properties of these other massive companions. Constraining formation models with period distribution observations is hence crucial, but becoming increasingly possible now with new data influx. \\

The variation in the period distribution of giant planets is first 
noticed in radial velocity surveys. \cite{udry2003} first mentioned a 
period valley between 10 and 100 days. This period valley sits between the hot Jupiters (HJ)\footnote{Planets on orbits shorter than 10 days.}  pile up at short periods (3-4 days) and warm Jupiters (WJs) beyond 100 days. We note that WJs are defined here as giant planets orbiting on periods between 10 and 300 days, slightly beyond the common definition ending at 100 days.
 A similar period distribution is seen in transiting giant 
planets observed by the {\it Kepler} Mission, although the strength of 
the HJ pile up may differs slightly from those from the radial velocity 
surveys \citep{howard2, dawson}. One difficulty of estimating 
the occurrence rate of giant planets from {\it Kepler} is the relatively large 
false positive rate. \citet{santerne2016} combined ground-based radial 
velocity follow up results with a magnitude-limited sample of giant planets 
discovered by {\it Kepler} and reported the occurence distribution of 
giant planets with orbital period smaller than 400 days around FGK stars. 
They found a HJ occurrence rate about half of 
what is found by the radial velocity surveys \citep{marcy2005, wright2012}, 
and confirmed a similar deficit of planets outside the period valley starting at 10 days orbits.

\begin{figure*}
\begin{center}

\includegraphics[scale=0.30]{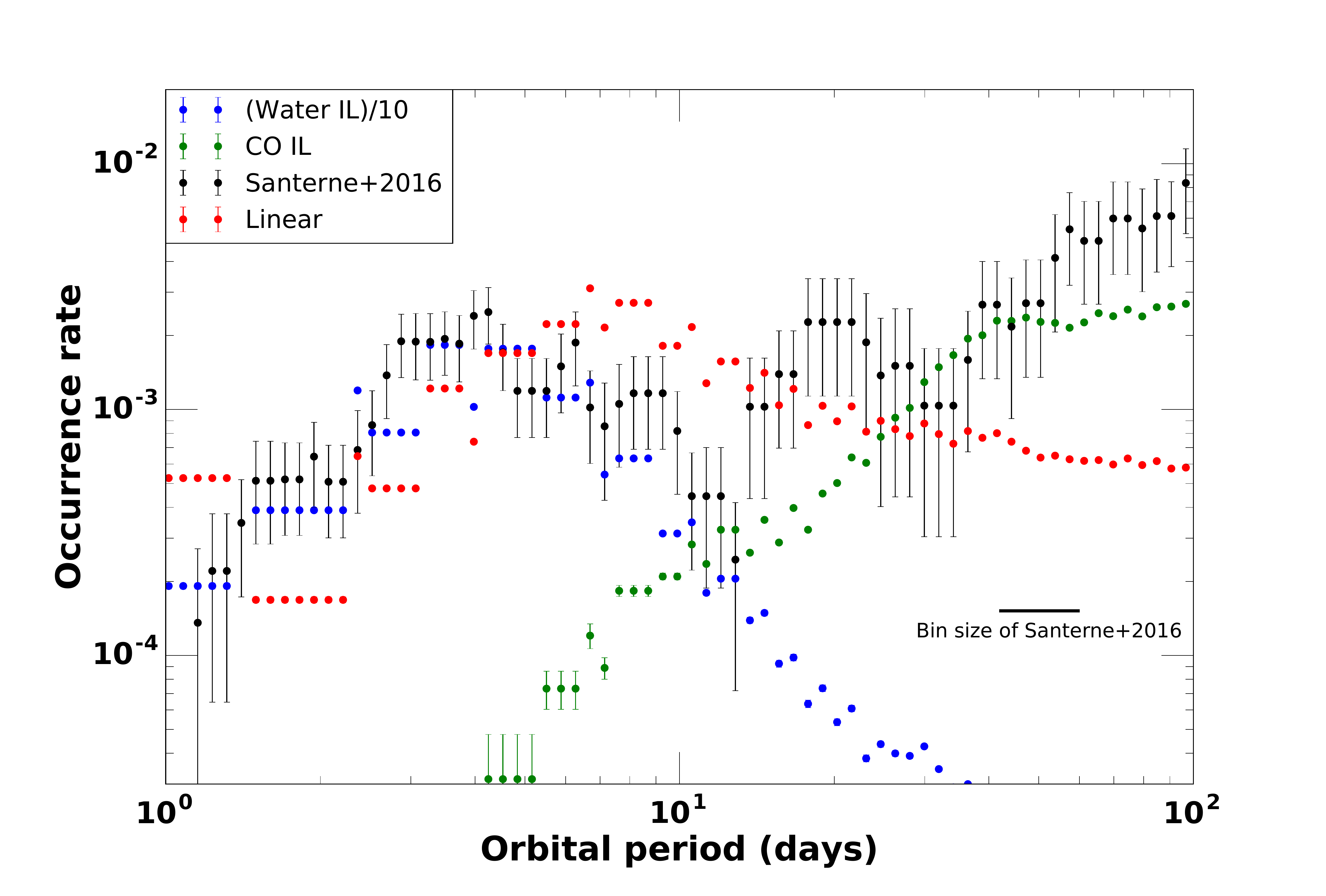}
\caption{The occurrence rate (per star) of giant (Jovian) planets as a function of their semi-major axis from \citep{santerne2016}. Note that to resolve the dip around 10 day period better, we plot the occurrence rate in a sliding bin, in which each data point and its error bars represent the occurrence rate and uncertainties in a bin centered at this data point in logarithmic space, and with bin width of 0.2333. We fit this distribution with two planet formation models. In the linear model, planetary seeds are injected randomly throughout the disk. In the bimodal model, seeds are injected solely at the water and CO icelines positions. The linear case leads to a near-constant occurrence rate of gas-giants, while the bimodal case is more compatible with observation. The bimodal case moreover predict a WJ/HJ ratio significantly closer to observations.}
    \label{fig:uncorrected}
    \end{center}
\end{figure*}

A summary of \cite{santerne2016} results is plotted in Fig. \ref{fig:uncorrected} showing the occurrence rate of the different giant planets classes as a function of the orbital period. We notice mainly the HJs pile up at 3 days and the dip in the occurrence rate around 10 days, where HJs end an WJs start. In total, WJs outnumber the HJs population significantly.\\ {Classical population synthesis models \citep{idalin,morda} were successful in reproducing multiple aspects of exoplanets statistics, for example the high occurrence rate of small planets, the planet-star metallicity correlation, and the low occurrence rate of intermediate mass planets. These models were however unable to reproduce the HJs pile up at 3 days, and attributed this to the migration scheme used \citep{benz2014,mordasini2015}. This pile-up however was reproduced by \cite{beauge} through high eccentricity migration of planets placed in systems with 3 or 4 planets starting at mean motion resonances. \cite{wu} on the other hand proposed that this pile-up can be explained by secular chaos in systems with 3 giant planets. All of these models however do not try to reproduce the he dip in the occurrence rate of gas giants at 10 days.}

{In this work, we offer an alternative explanation to the gas-giants occurrence rate profile by fitting it to a populations synthesis model based on pebble accretion.  The goal is to check multiple families of initial conditions and compare them to the observations. Specifically, we want to understand whether this population is recovered better from a stochastic linear distribution of planetary seeds, or if formation only at special locations in the disk is needed to retrieve the bimodal distribution seen in Fig. \ref{fig:uncorrected}. Our model has the advantage of recovering the occurrence rate details entirely through disk migration.}\\


\section{Model}
The model we use in this work is based on \cite{ali-dibb,ali-dibc}, following \cite{lamb1,lamb3,bitsch1}, \cite{bitsch2,morby1}. It includes the following:
\begin{itemize}
\item Fits to a radiative 2D disk model with accurate opacities transitions leading to structures in the disk. {We note however that since these simulations were done for a constant 1 solar mass star, and a constant turbulent $\alpha \sim 5\times 10^{-3}$, we do not vary these parameters to be consistent with the simulations.}
\item Parametric pebbles and gas accretion including pebble accretion in both the Bondi and Hill regimes, in addition to slow and fast phases gas accretion.
\item Type I and II migration through torque evaluation. Type I migration will affect low mass planets through the Lindblad and corotation torques, while type II migration will affect planets massive enough to open a gap in disk and follow its viscous evolution. We assume that the planets inward migration will stop at the inner cavity, and hence will not be lost to the star. {It is however important to note that we find no inner grid boundary pile-up of HJs, and thus the inner boundary condition have no effect on our model. Moreover, our inner visualization bin starts at 0.7 days (beyond the inner edge of the grid), to be consistent with \cite{santerne2016}. }
\item Photoevaporation (PE) can increase the metallicity of the disk disk and thus affect its opacity. We assume a simplistic PE model, where we modify the accretion rate controlling the disk structure by reducing the PE mass flux from it till eventually it reaches 0 where the disk is assumed to be completely dispersed. Photoevaporation will remove the disk's gas while retaining the dust, leading to gradual increase in its metallicity, which we integrate into the model \citep{guillothueso}. We however do not take into account the viscous spreading of the disk due to PE. We are hence replacing the disk global accretion rate $\dot{M}_{acc}$ of \cite{bitsch2} by:
\begin{equation}
\dot{M}'=\dot{M}_{acc}-\dot{M}_{PE}
\end{equation}
and then define the disk's gas metallicity enhancement as:
\begin{equation}
    \varepsilon_c= 1 + \frac{\dot{M}_{PE}}{\dot{M}_{acc}}
\end{equation}

\item Simulations are stopped when either the disk fully dispersed, or when the planet reach the inner edge of our disk at 0.01 AU. 
\end{itemize}
Moreover, we modified the model above to take into account the growth of small planetary seeds. In the earlier models we injected seeds with masses $\sim10^{-4}$ M$_{\oplus}$, close to the pebble transition mass, and their growth was dominated by pebble accretion. In this work, however, we start with smaller seeds with masses = $10^{-5}$ M$_{\oplus}$ (corresponding the a radius of 160 km, in the same order of the observed bump in the asteroids size distribution \citep{asteroids}), and hence we self-consistently incorporated the relevant weak coupling branch into the model. We hence follow \cite{johansen2015, johansen2017} by defining the effective accretion radius in the Bondi regime as:
\begin{equation}
\hat{R}_{acc}=\bigg( {\frac{4 \tau_f}{t_B}}\bigg) ^{1/2} \ R_B
\end{equation}
for the weak coupling branch ($\tau_f > t_B$ and $R_B < R_H$) we follow \cite{ormel2010} in modifying this accretion radius as :
\begin{equation}
\hat{R}_{acc}=\hat{R}_{acc}\times\exp(-0.4\times (\tau_f/t_p)^{0.65})
\end{equation}
with the characteristic passing time-scale:
\begin{equation}
t_p=GM/(\Delta v + \Omega R_H)^3
\end{equation}
Moreover, we also take into account planetesimal accretion that is important for seeds in this mass range, specially in the inner disk. We hence follow \citep{bitsch2} in defining the corresponding accretion rate as:
\begin{equation}
\dot{M}_{c,plan}=3\times 10^{-4} \bigg(\frac{10 AU}{r_p}\bigg)  R_H v_H\Sigma_{peb}
\end{equation}
where $v_h$ is the Hill velocity and $\Sigma_{peb}$ the pebble surface density.

Such a global model includes a large number of free parameters. To keep the problem tractable, we only vary the parameters that are assumed to affect directly the planets occurrence rates, shown in table \ref{t1}. The free parameter space is explored through a population synthesis approach. The seed injection time (T$_{\text{ini}}$) is drawn linearly, while the seed injection location ($R_0$) is drawn either linearly or bimodally (snowlines). The dust metallicity (in small coupled dust grains) and Z$_0$ (the pebble metallicity) on the other hand is drawn from a Gaussian distribution with the mean and standard deviation of the stars sample used in \cite{santerne2016}. The disk's FUV photoevaporative flux ($\dot{M}_{FUV}$) is also drawn from a Gaussian distribution reflecting the disk age distribution of \cite{hernandez}. The rest of the problem's free parameters are assumed to be constant, including M$_0$ (the seed's initial mass), $f$ (a fudge factor that reconciles our simplified slow phase gas accretion rate parametric fit with more detailed hydrodynamic simulations), $\kappa_{env}$ (the envelope opacity), and $\rho_c$ (the core's density). These parameters are explained more in detail in \citep{bitsch2, ali-dibb}.

\subsection{Dynamical properties}
The main caveat in this model is not taking into account the dynamical evolution of planets, even though half of the WJs in the RV sample (\cite{wright} have significant eccentricities ($e\gtrsim 0.2$ , and cf. The \texttt{exoplanets.org} database\footnote{Consulted on February 1st 2017.}). This is problematic because disc-planet interactions are not expected to excite large eccentricities \citep{bitsch2013}. Moreover, it is not clear why these planets have parked on these orbits instead of migrating all the way to become HJs.\\

On the other hand, even though eccentricities can be excited by planet-planet scattering, at small enough semi-major axes ($a\lesssim0.5$ AU for a Jupiter-like planet) this will lead to planet-planet collision with small eccentricity excitation ($e\lesssim0.1$) \citep{ford2001, johansen2012, petrovich2014}. 
One possible solution is planet-planet scattering during early dynamical instabilities \citep{lega2013, crida2016}. Another possibility is based on the intriguing trend that WJs with outer planetary companions have a significantly wider eccentricity distribution than the sample without companions \citep{dong2014,petro2016}. This sample could have undergone high-eccentricity migration (through Kozai oscillations followed by tidal circularization) \citep{dawson2014}. However, tides are too weak at these relatively wide orbits be effective.

\cite{petro2016} proposed that this population is transient, where the planets are undergoing continuous migration from secular planet-planet or star-planet interactions, and we only observe them at the low eccentricity phase of this migration, and showed that such mechanism can reproduce their eccentricity distribution. Therefore, a fraction of WJs with the largest eccentricities ($e\gtrsim0.4$) migrate through this mechanism, while the fraction with lower eccentricities migrate through another channel. \\

Another dynamical property of gas-giants is their spin-orbit alignment (the angle between their orbital axis and the spin axis of their parent stars). However, there is virtually no constraints on the WJ population from spin-orbit angles. Although many of them are in multiple transiting planet systems \citep{huang2016}, which are expected to be aligned with their host star. A notable exception is HD80606b \citep{winn2009} with 45 deg angle. For HJs, statistics from the \texttt{exoplanets.org} database show a median absolute angle of 13.8 deg for this population. Therefore roughly 50\% of HJs have spin-orbit misalignment. {\cite{cridabat} however concluded that the spin-orbit misalignment of HJs is compatible these having been transported via disk migration in a disk torqued by a companion. In this case, both aligned and misaligned HJs could have formed on the snowline and then disk-migrated inward as per our model, explaining the pile-up observed in both populations \citep{winnfab}. }


\section{Results \& Discussions}
\subsection{Analytical considerations}
\label{analytic}
The gas giants occurrence rate profile in Fig.\ref{fig:uncorrected} is spread out over two orders of magnitude, and appears to be bimodal with a bell-like distribution inside 10 days and a power-law beyond it. It is hard to imagine how to get such structure using a classical protoplanetary disk model with stochastic initial distribution of planetary seeds. 

Let us assume a basic protoplanetary disk where temperature and density follow simple power-laws. The solid accretion rate onto a core and its disk type I migration speed both scale linearly to the disk's density. Therefore, a random initial distribution of planetary embryos will lead to a near-constant final distribution of gas giants. In other words, if we inject enough planetary seeds (while exploring the entire free-parameter space) in the disk, we expect the resulting population of gas giants to occupy every possible final location, since all of the processes in this toy model are linear.


A possible way to generate this bimodal distribution is if planets form preferentially at specific locations in the disk. The most interesting permanent disk structures to consider are the main volatiles condensation fronts (snowlines). This is motivated theoretically by multiple works arguing that snowlines can be preferred places for planets formation \citep{ros,ali-diba,idaguillot,schooormel}, and observationally by the radial gaps seen in TW Hya \citep{twhya1,twhya2} and HL Tau \citep{hltau}, and their correlation with the positions of icelines \citep{zhang}.   

Since these are fundamentally temperature-dependent, their location will vary within the same disk with time as it cools down. If planet seeds form preferentially at two snowlines (water and CO for example), then even with a completely linear disk we might end up with a bimodal distribution.

\subsection{Simulations}
We first run simulations with linear initial distribution of planetary seeds (as shown in table \ref{t1}). Resulting occurrence rates as a function of period are presented in Fig. \ref{fig:uncorrected}. This result conforms to what we expected in the analytical discussions, which is a quasi-linear final distribution of gaseous giant. The small bump inside 10 days can be attributed to {type I migration. It is analogues to the over abundance in HJs found in the classical population synthesis models. This was attributed to short type I migration timescale leading to a big pile up of planets at the inner edge of the disk. Since our model incorporates the corotation torque, slowing down type I migration, in addition to the fast pebbles accretion (decreasing the time a planet will take to open a gap), the huge edge of the grid pile-up of classical population synthesis models translates into the mild pileup at 8 days.}


We now run simulations assuming that small planetary seeds form preferentially on the water and CO icelines. We hence inject the seeds exclusively at the (evolving) snowline positions, calculated via the disk model. The younger a disk is, the hotter it is and thus the farther the snowlines are. This will lead naturally to a bell-like occurrence rate for each snowline, resembling that seen in Fig. \ref{fig:uncorrected}. This however works only if we give the two icelines different weights by reducing the planets formation efficiency rate of the water iceline by a factor between 3 and 10. We are hence assuming that either the water iceline forms planetary embryos less efficiently than we assumed, or that a significant fraction of its planets are lost to the star \citep{lost1,lost2} \\

The main result from our simulations is that a linear distribution of planetary embryos will lead to a quasi-linear final distribution of gas-giants, while a bimodal distribution of seeds (on snowlines) will lead to 2 clusters. To understand more the physical origin of this let us consider the following simple case. 

Let us fix the seed injection location for a planet to a specific radius in the disk, for example 10 AU. The growth/migration track of this planet will depend on the disk temperature/density structure around and inside of 10 AU. This disk structure is time dependent, so planets forming at 10 AU at different times will encounter different disk structures and thus follow different growth tracks. Therefore, our 10 AU seed, injected at different times in the disk will end up at different locations. If we integrate this over all possible starting locations and disk free-parameters, the resulting gas-giants will occupy every possible final location in the disk and thus lead to a quasi-linear occurrence rate profile.

\begin{figure}\includegraphics[scale=0.25]{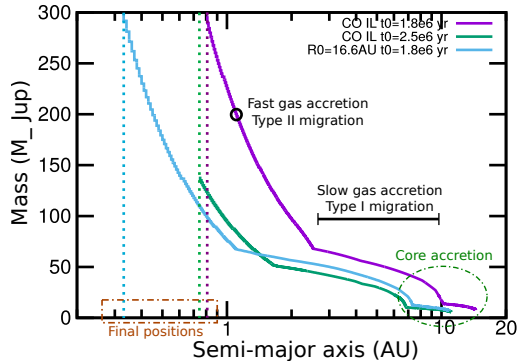}
\begin{center}
   \caption{The growth tracks for multiple planets injected at different times and locations in the disk. The two tracks for the seeds injected at the CO iceline at different times converge to roughly the same location since they encountered similar disk density and temperature profiles due to starting at the same temperature. The seed injected at the same location where the iceline was but at a different time end up relatively far from the other two cases, since it encountered a different disk structure due to it starting at a different temperature.}
    \label{fig:track}
    \end{center}
\end{figure}

On the other hand let us imagine planetary seeds placed exclusively on a snowline. Since the snowline is a point in the temperature/density profile of the disk and not a fixed radius, planets forming at this point at different times will experience roughly similar density/temperature profiles inside their location and thus their formation tracks will converge around a specific location leading to clustering. This can been seen in Fig. \ref{fig:track}. \\

{We note that mixing the linear model with the CO iceline planets will lead to an occurrence rate profile that resemble somehow observations, but shifted to the right. This hence will fit neither the Hot-Jupiters pile up or the dip at 10 days. It is however hard (if not impossible) to tweak the parameters in a way that makes this work. This is because the CO iceline planets will always have the same distribution controlled by the CO condensation temperature which is not a parameter. Thus the only degree of freedom is the linear case planets. To push this distribution left we need a cut-off in the possible initial location of planets at some orbital period. This seem unnatural within the physics included in the model. Moreover, it is not clear why their would be this cutoff, only to be followed further out by a very active CO iceline. }

It is important to mention that all of our simulations are scalelable vertically, meaning that, assuming statistical significance, we are allowed to multiply our occurrence rates by a constant value for the entire simulation. This is because we are trying to fit the relative occurrences rates of the different planetary populations, not the absolute abundance of gas-giants. 

\subsection{The effects of the model's parameters}

\begin{figure*}
\begin{center}
\includegraphics[scale=0.30]{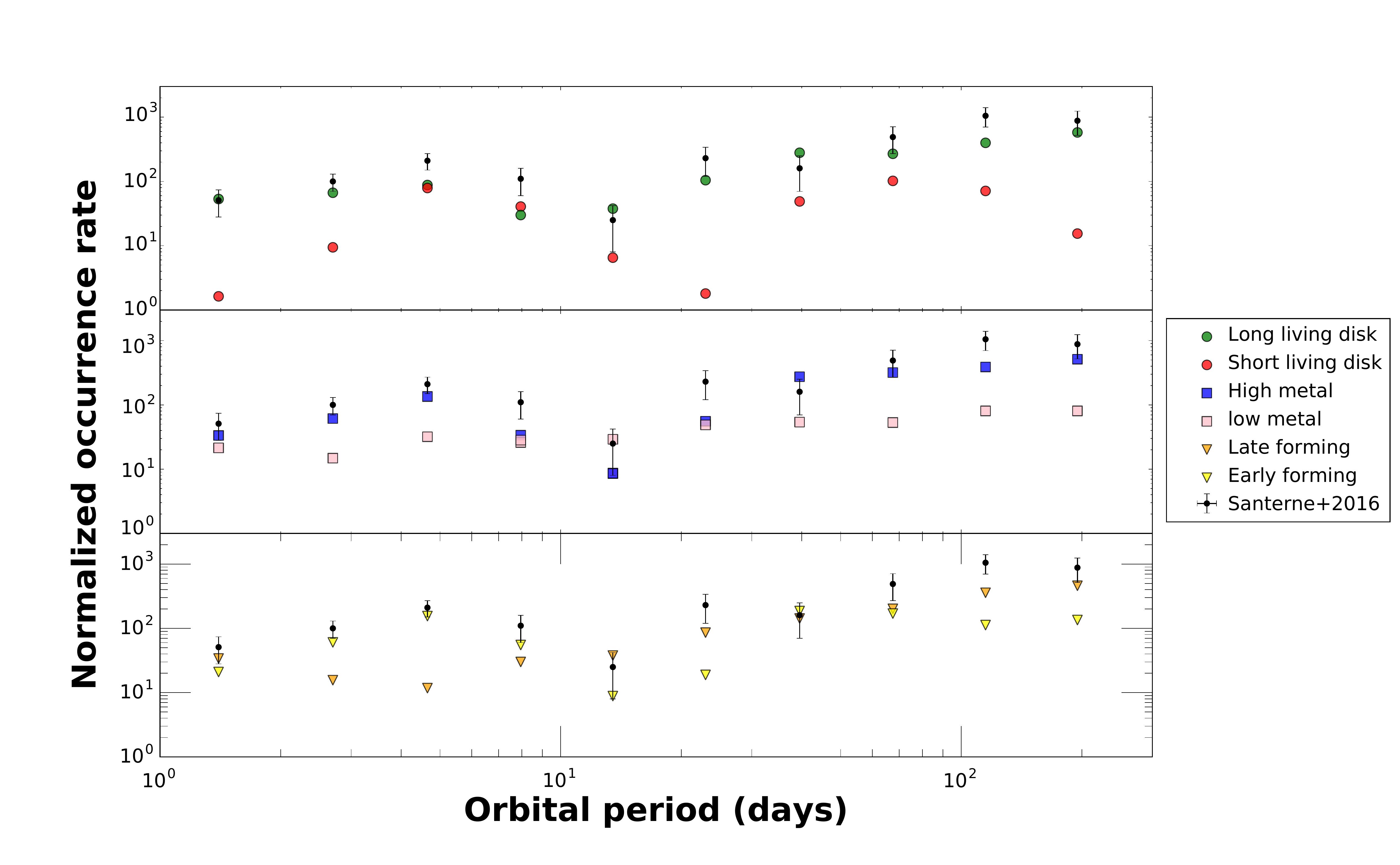}
   \caption{Effects of the different model parameters on the occurrence rate of gas giants. Green and red circles represent models with $\dot{M}_{FUV}$ respectively less and more than 2$\times 10^{-9}$ M$_{\odot}$/yr. Blue and pink squares represent models with disk grain metallicity respectively higher or less than 0.47 \%. Orange and yellow triangles correspond to models with planets seed injection times of respectively more and less than 2 Myr. We notice that the occurrence rate is higher for planets either forming early, or in long living disks, or in disks with high metallicity. This is expected since these conditions favor the formation of gas giants who need enough metals and a lot of time to form. } 
    \label{fig:param}
    \end{center}

\end{figure*}

\begin{figure}
\begin{center}
\includegraphics[scale=0.20]{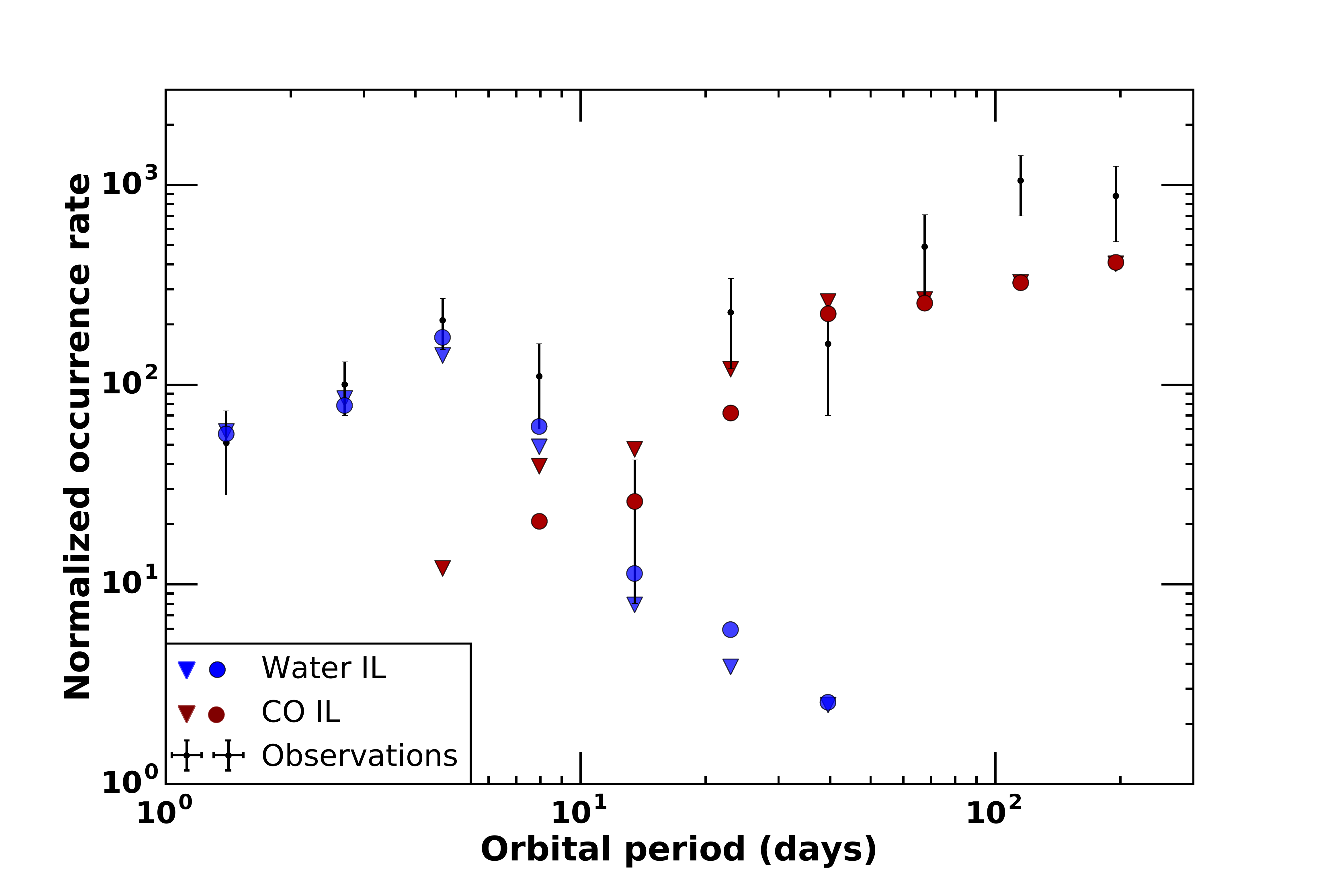}
   \caption{The effect of the envelope opacity on the occurrence rate of gas giants. Circles represent our nominal opacity case (0.02 cm$^2$ g$^{-1}$) and triangles represent higher opacity (0.05 cm$^2$ g$^{-1}$). The differences between the two cases are minimal.} 
    \label{fig:newop}
    \end{center}

\end{figure}

{To better understand the effect of the different parameters explored in the population synthesis, we split the range used for each parameter into two halves at the median value, and visualize the occurrence rate for each of them.
The photoevaporative mass flux will affect a disk's dispersal time and metallicity. Disks with lower PE mass fluxes will live longer, thus giving more time for giant planets to form. We hence expect this parameter to affect the giant planets distribution by increasing the occurrence rates for lower fluxes. The disk metallicity on the other hand controls the amount of solids available for planets formation. Moreover, it affects the disk structure through opacity. To first order, due the planet-star metallicity correlation \citep{fischer,guillot} we expect disks with higher metallicities to be more efficient at forming planets. Results of parameters exploration are shown in Fig. \ref{fig:param}. We notice that the occurrence rate of gas giants is dominated by high metallicity and long living disks. This is not surprising since these parameters give a gas giant enough solid materials and time to form. The effect of when did a planet start forming in the disk (early vs late) is less trivial, since forming early will give a planet more time to evolve into a gas giant, but also will affect where it is going to end up in the disk. This non linear effect is the reason why CO iceline planets are dominated by planets forming late in the disk.
Interestingly, we notice that the overall occurrence rate shape (width and depth) is robust to the explored parameters ranges. This indicate that this shape is controlled by the underlying physical model and its implicit assumptions (accretion and migration speeds, disk model), rather than by our choice of parameters. Other -fixed- parameters (icelines temperatures and core density) are invariable physical quantities that will not differ between systems. The effect of the only remaining parameter, envelope opacity, is shown in Fig. \ref{fig:newop}. This plot compare the occurrence rates for identical models with 2 different envelope opacity parameter: 0.02 and 0.05 cm$^2$ g$^{-1}$. The effect of this change is minimal, with the depth and width of the profile unchanged for both water and CO iceline planets. }





\subsection{WJ/HJ ratio}
A more general and bin-size independent method of comparing the models to data is through the WJs to HJs occurrence rates ratio (W/H). This should give a basic but solid information on the accuracy of the models in reproducing the relative abundances of the two giant planets population. We hence calculate this ratio (where HJs are inside 10 days and WJs are beyond this) for the observational data and the two models. From \cite{santerne2016}, the data W/H is $\sim$ 8.3.
The linear model on the other hand gives a W/H of $\sim$ 1. This near-unity value implies that the linear model predicts as many HJs as WJs, which is expected from the analytic considerations, and from the fact that even though the WJs space is larger, the inner disk (translating into HJs) is more efficient at forming planets.

The icelines model gives a  W/H $\sim$ 8 only after decreasing the efficiency of the water iceline by a factor 10. The icelines case provides fit better than linear case for water iceline efficiency ranging from 1.25 (where it leads to W/H=1) to 10, where it matches observations.

This moreover can be improved if we assume that the 50$\%$ of WJs with high eccentricities all formed via dynamical instead of disk migration. This therefore can allow us to decrease the data W/H to $\sim$ 4, and thus fit the data perfectly by reducing the water iceline efficiency by a factor 5. {This however does not take into effect Hot Jupiters who might have reached their current orbits via high eccentricity migration followed by tidal circularisation. If this population is significant, then this will increase the measured W/H ratio back to near 8.}

This implies that our model either overestimates the abundances of HJs, or underestimate the abundance of WJs. In the first case scenario our model would be similar to the earlier population synthesis models that predicted a pileup of HJs due to type I migration. In the second case scenario, an additional source of WJs might be needed. Other structures in the disk can possibly play this role. For example, the N$_2$ iceline should be close to the CO iceline since the two elements condense at comparable temperatures \citep{fray}. A significant fraction of planets forming at this location should therefore end up as WJs, in parallel with the CO iceline planets. Another possible location is the outer edge of the deadzone, where the viscosity transition can trigger a Rossby wave instability \citep{lyra} leading to an accumulation of solids that might trigger planets formation.

\subsection{Statistical analysis}
To test the statistical significance of our findings we conduct a Gaussian Mixture Model (GMM) analysis that predicts the optimal number of Gaussian components that fit the data and the simulations. Models with low Bayesian information criteria (BIC) value are preferred to those with higher values \citep{hastie}.

\begin{figure}
\begin{center}
\includegraphics[scale=0.35]{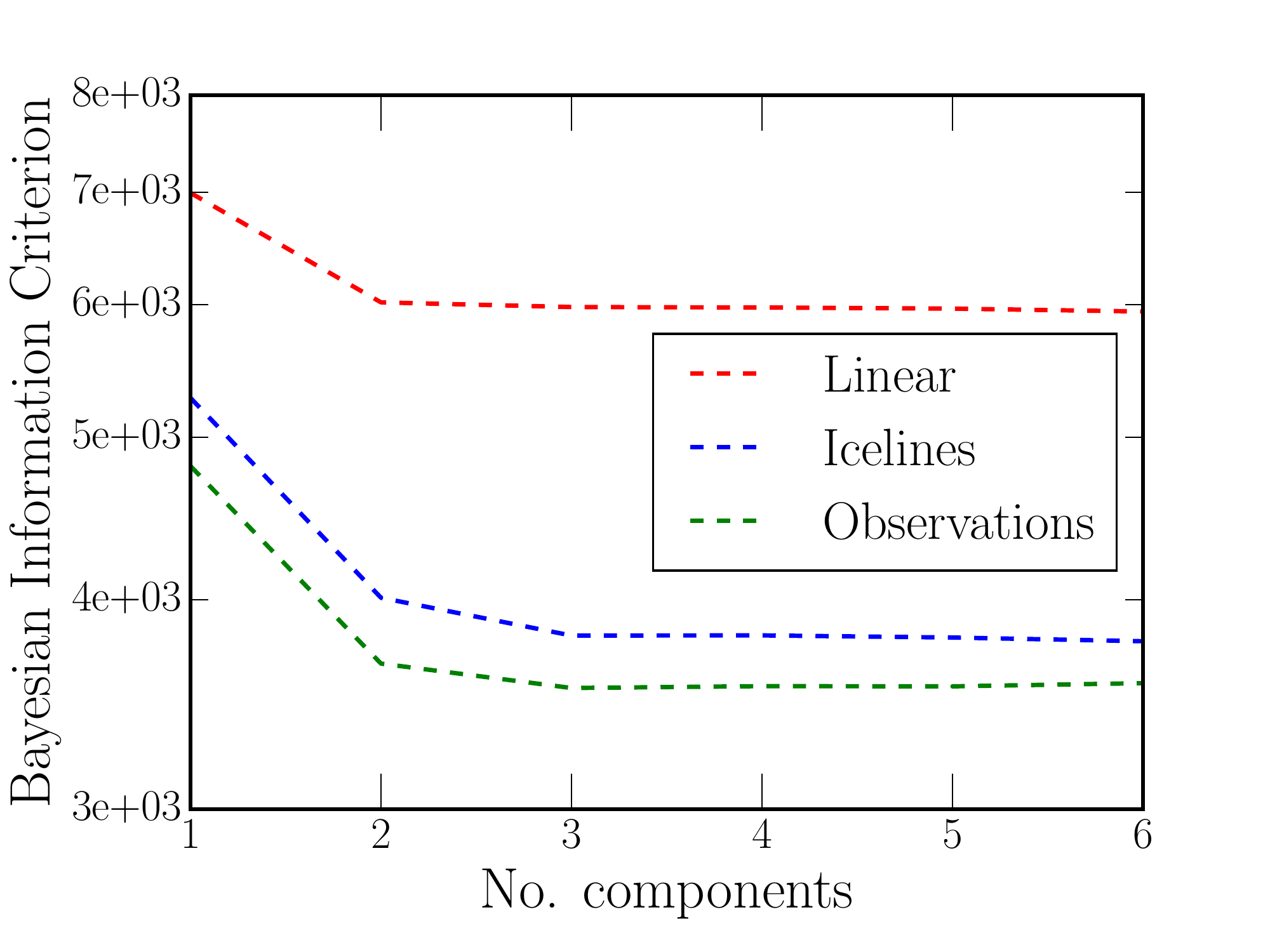}
   \caption{Gaussian Mixture Model analysis for the observational data and two models. {We notice that the information gain (BIC decrease) from 1 to 2 components model is greater for the observations and icelines model than for the linear model. This implies that the observations and icelines model are both more strongly bimodal than the linear case.}}
    \label{fig:chelsea}
    \end{center}

\end{figure}

Our results in Fig. \ref{fig:chelsea} show that the observational data significantly favor two Gaussian components over one, and so does both the Icelines and linear cases. {However, the icelines case have a steeper slope between 1 and 2 components than the linear case, implying that it prefers 2 components more strongly than the linear case, thus favoring it as a fit to the data.}
 
{The ratio between the BIC score for a two/multi-component GMM model and a one
component GMM model tells us about the significance of how bimodal/multimodal the
data is. In our particular case, the iceline model and observational data
are both more strongly bimodal when compared to the linear model, because the slope of their BIC is steeper between 1 and 2 components. This is different from
the standard KS tests because these are most sensitive when the underline distributions
differ in a global fashion near the center of the distribution. However, it is possible to make
centers of distribution similar between a single mode and a bimodal distribution. Since we are more 
interested in if the giant population is bimodal, a BIC test with GMM model is more appropriate compared to a KS test. }

\subsection{Predictions}
\begin{figure}
\begin{center}
\includegraphics[scale=0.20]{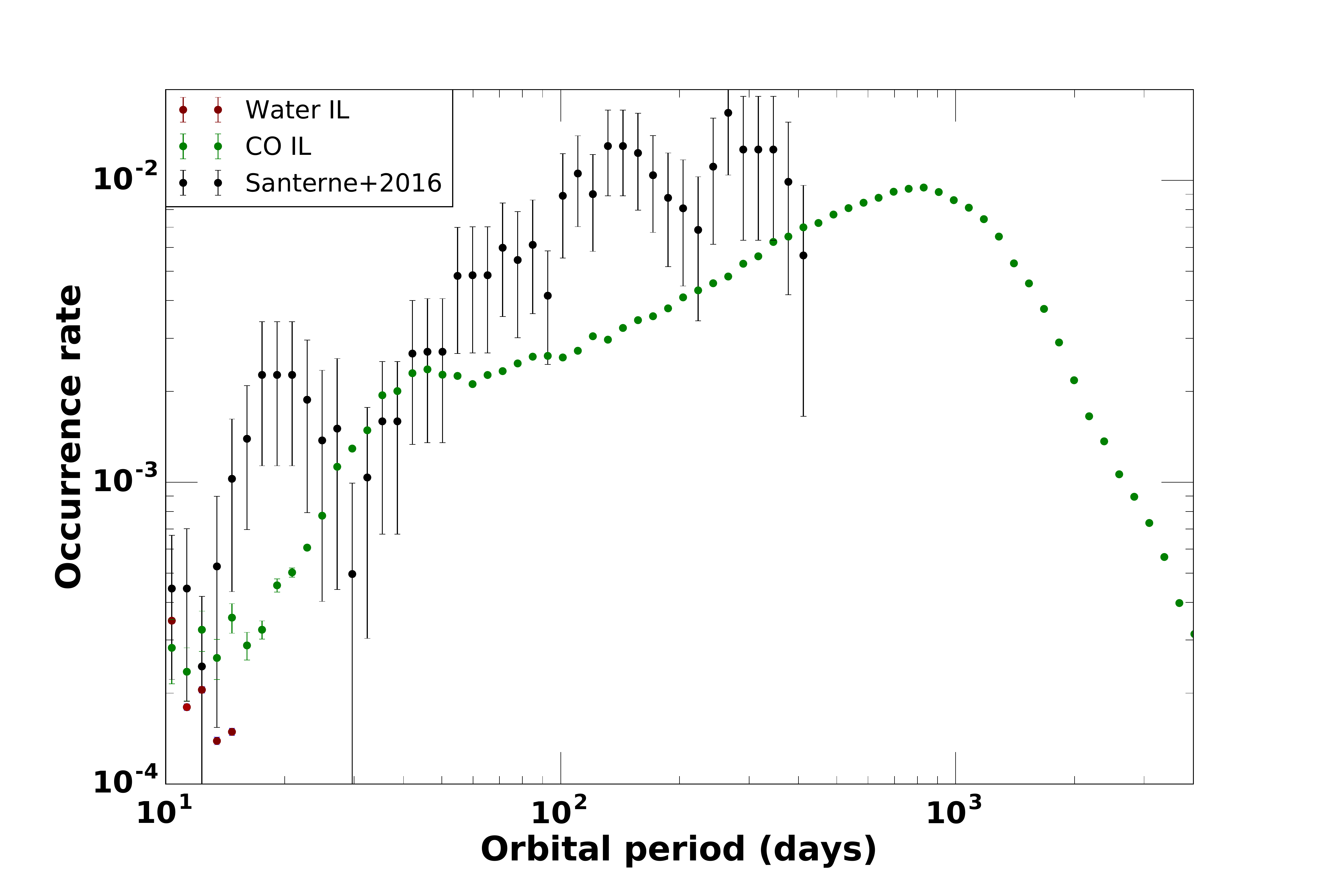}
   \caption{Predictions from our model. This plot shows the occurrence rate of gas giants beyond 100 days as predicted by our icelines model. These planets all started forming at the CO iceline. We predict that, similar to the water iceline planets, the CO iceline planets will follow a bell-like occurrence rate profile, with a central pile up and smooth decrease on both sides.}
    \label{fig:pred}
    \end{center}

\end{figure}

{Our predictions from this model are show is Fig. \ref{fig:pred}. Since in our model planets form exclusively on the water and CO icelines, and since both follow the same physics, we expect the occurrence rates of gas giants generated by the 2 icelines to follow similar profiles. This is validated by the simulations, where CO iceline gas giants follow a bell-like distribution with a central pile-up. We predict this pile up to be no further than 1000 days orbits, followed by a steady decline.}

\section{Summary \& Conclusions}
In this work we investigated the origin of the occurrence rate radial profile of gas-giants inside 300 days found by \cite{santerne2016}. We used a population synthesis model based on pebble accretion including solids and gas accretion, disk migration, and simplified photoevaporation to fit the observational data. Starting from a linear distribution of planetary seeds throughout the disk, our simulations produce a quasi-linear final distribution of planets with a near unity WJ/HJ ratio, and thus fail to properly fit the data. If we inject planetary seeds solely on the water and CO icelines however, we get a much better statistical fit, assuming a factor 3-10 lower efficiency for the water IL. Moreover, we conducted a Gaussian mixture model analysis showing that the icelines model have a lower BIC score than the linear model, indicating that it is a better fit to data. Our results exclude simple models with linear initial distribution of planetary seeds, and hint toward snowlines being preferred places for planets formation. {Our model can be improved on multiple fronts. The most significant missing element is planets dynamics. In this model we use disk migration to move planets forming on icelines inward to where they are observed. We however do not see any fundamental reason why these planets cant form at the icelines and then migrate dynamically inward via Kozai/scattering/secular migration, thus explaining the eccentricities of WJs. These are highly non-linear effect that needs detailed modeling. Another possible relevant effect is snowline fossilization \citep{morbyfos} that becomes important when forming multiple gas giants in a single disk. }

\renewcommand\arraystretch{1.0}
\begin{table}
\begin{center}
\caption{Initial conditions}
\tiny
{\begin{tabular}{lcc}
\hline
\noalign{\smallskip}

Linear parameters			& Range		\\

\hline
T$_{ini}$			& 10$^5$ yr - disk dissipation 		 \\
R$_0$	& 0.5 - 30 AU						 \\	
\hline

Gaussian distributions			& $\mu$			& $\sigma$	\\
\hline
metal (\%) & 0.47 & 0.7  \\
$\dot{M}_{FUV}$ (M$_{\odot}$/yr) & $2 \times 10^{-9}$ & $2 \times 10^{-9}$ \\

\hline
M$_0$	(M$_\oplus$)			&  10$^{-5}$	& -		\\	
Z$_0$ (\%) & 2 \ $\times$ \ metal & -  \\
$f$ & 0.2 & - \\
$\kappa_{env}$ (cm$^2$ g$^{-1}$) &  0.02 	& - \\
$\rho_c$ (g cm$^{-3}$) & 5.5 & - \\
\hline		
H$_2$O {iceline} & 150 K & -  \\
CO {iceline} & 25 K & -  \\
\hline

\end{tabular}}\\

\label{t1}
\end{center}
\end{table}



\section*{Acknowledgements}

We thank an anonymous referee for useful comments that significantly improved this manuscript. We thank C. Petrovich for reading and commenting on this manuscript. AJ acknowledges the support from the Knut and Alice Wallenberg Foundation (grants 2012.0150, 2014.0017, 2014.0048), the Swedish Research Council (grant 2014-5775), and the European Research Council (Starting Grant 278675-PEBBLE2PLANET). Special thanks go to the Centre for Planetary Sciences group at the University of Toronto for useful discussions.








\appendix


\bsp	
\label{lastpage}
\end{document}